\documentclass[letterpaper]{article} 
\usepackage{aaai25}  
\usepackage{times}  
\usepackage{helvet}  
\usepackage{courier}  
\usepackage[hyphens]{url}  
\usepackage{graphicx} 
\usepackage{natbib}  
\usepackage{caption} 
\usepackage{algorithm}
\usepackage{algorithmic}

\usepackage{multirow}
\usepackage{subcaption}
\usepackage{mathtools}
\usepackage{booktabs}       
\usepackage{amsfonts}       
\usepackage{nicefrac}       
\usepackage{microtype}      
\usepackage[table]{xcolor}
\usepackage{newfloat}
\usepackage{listings}
\DeclareCaptionStyle{ruled}{labelfont=normalfont,labelsep=colon,strut=off} 
\floatstyle{ruled}
\newfloat{listing}{tb}{lst}{}
\floatname{listing}{Listing}
\title{Multi-Modal Latent Variables for Cross-individual Primary Visual Cortex Modeling and Analysis}
\author{
    Yu Zhu\textsuperscript{\rm 1,2}\footnote{This work was done during his internship at Shanghai AI Lab},
    Bo Lei\textsuperscript{\rm 4},
    Chunfeng Song\textsuperscript{\rm 3}\footnote{ Corresponding Author},
    Wanli Ouyang\textsuperscript{\rm 3},
    Shan Yu\textsuperscript{\rm 1}$^\dagger$,
    Tiejun Huang\textsuperscript{\rm 4}
}
\affiliations{
    \textsuperscript{\rm 1}Institute of Automation, Chinese Academy of Sciences\\
    \textsuperscript{\rm 2}School of Artifcial Intelligence, University of Chinese Academy of Sciences\\
    \textsuperscript{\rm 3}Shanghai Artificial Intelligence Laboratory\\
    \textsuperscript{\rm 4}Beijing Academy of Artificial Intelligence\\
    zhuyu2022@ia.ac.cn, songchunfeng@pjlab.org.cn, shan.yu@nlpr.ia.ac.cn 
}

\begin{document}

\maketitle

\begin{abstract}
Elucidating the functional mechanisms of the primary visual cortex (V1) remains a fundamental challenge in systems neuroscience. Current computational models face two critical limitations, namely the challenge of cross-modal integration between partial neural recordings and complex visual stimuli, and the inherent variability in neural characteristics across individuals, including differences in neuron populations and firing patterns. To address these challenges, we present a multi-modal identifiable variational autoencoder (miVAE) that employs a two-level disentanglement strategy to map neural activity and visual stimuli into a unified latent space. This framework enables robust identification of cross-modal correlations through refined latent space modeling. We complement this with a novel score-based attribution analysis that traces latent variables back to their origins in the source data space. Evaluation on a large-scale mouse V1 dataset demonstrates that our method achieves state-of-the-art performance in cross-individual latent representation and alignment, without requiring subject-specific fine-tuning, and exhibits improved performance with increasing data size. Significantly, our attribution algorithm successfully identifies distinct neuronal subpopulations characterized by unique temporal patterns and stimulus discrimination properties, while simultaneously revealing stimulus regions that show specific sensitivity to edge features and luminance variations. This scalable framework offers promising applications not only for advancing V1 research but also for broader investigations in neuroscience.
\end{abstract}

\section{Introduction}
The primary visual cortex (V1) plays a fundamental role in hierarchical visual information processing, making its functional characterization essential to systems neuroscience. Recent advances in calcium imaging technology \cite{sofroniew2016large} have accelerated V1 research by enabling recording of large neuronal populations in multiple subjects. A central goal in utilizing this technology is to identify distinct patterns of neural activity and establish their relationships with specific visual stimuli \cite{stringer2019high}.

\begin{figure}[t]
    \centering
    \includegraphics[scale=0.57]{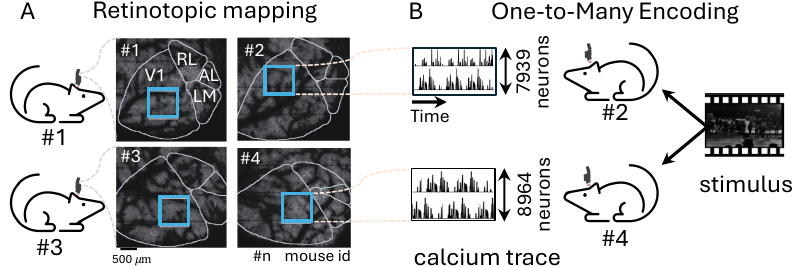}
    \caption{\textbf{Challenges in modeling V1.} (A) Retinotopic mapping (RL, AL and LM refers to other visual cortex) and the partially recorded field of view of calcium imaging (blue rectangle). (B) One-to-many challenge with cross-individual heterogeneity, especially neuron counts and firing patterns.}
    \label{intro-challenges}
\end{figure}

Current approaches to understanding V1 visual information processing primarily follow two paradigms. The first develops encoding models \cite{sinz2018stimulus,wang2023towards} that map visual stimuli to neural activity, simulating V1 function. The second focuses on decoding neural activity to understand the representation of visual features such as edges and motion \cite{yoshida2020natural}. However, these approaches face fundamental limitations due to the restricted shared subspace between sparse, locally-recorded neural activity and complex visual features, making it challenging to fully capture brain-vision relationships. Furthermore, these models often make the oversimplified assumption that locally recorded V1 populations can fully encode complete visual stimuli, despite evidence that visual processing is distributed across the extensive V1 area according to retinotopic mapping (Figure \ref{intro-challenges}.A).

Additionally, cross-individual neural heterogeneity presents an additional significant challenge. Neural responses to identical stimuli can vary substantially among individuals \cite{guntupalli2016model,haxby2001distributed}, complicating cross-individual modeling efforts (Figure \ref{intro-challenges}.B). Addressing this variability requires zero-shot or few-shot domain adaptation capabilities, a challenge that remains significant even in contemporary machine learning. Previous approaches \cite{wang2023towards} requiring individual-specific fine-tuning have not adequately addressed this cross-individual variability.

To address these challenges, we introduce a \textbf{m}ulti-modal \textbf{i}dentifiable \textbf{v}ariational \textbf{a}uto\textbf{e}ncoder (miVAE) featuring two-level latent space disentanglement. For neural activity, we separately model idiosyncratic and preserved latent variables, capturing both subject-specific characteristics and functionally consistent information across individuals. For visual stimuli, we distinguish between neural activity-related and unrelated latent variables, refining the preserved variables to capture highly relevant neural correlates. We further enhance these highly correlated variables through latent modeling and introduce a novel score-based attribution strategy for comprehensive data interpretation. Validated on a mouse V1 dataset \cite{turishcheva2023dynamic} using two-photon calcium imaging \cite{grienberger2012imaging}, miVAE demonstrates superior cross-individual latent coding without requiring individual-specific fine-tuning, showing scalable performance with increasing dataset size and remarkable consistency across individuals exposed to identical visual stimuli. This cross-individual approach significantly enhances the scalability of data-driven research. Our attribution analysis successfully identifies key neuronal subpopulations with distinct stimulus-related responses and superior discriminative capabilities, while highlighting V1's sensitivity to edge and luminance patterns. This framework presents a powerful tool for V1 research with potential applications across various sensory cortices.

Our primary contributions include:
\begin{itemize}
    \item \textbf{Multi-modal identifiable modeling with latent space refinement.} miVAE effectively captures highly correlated variables between visual stimuli and neural responses, enhanced through pair-wise modeling in the latent space. 
    \item \textbf{Scalable cross-individual performance.} Our approach achieves state-of-the-art modeling and cross-individual alignment without subject-specific fine-tuning, demonstrating improved performance with increased data size.
    \item \textbf{Score-based attribution analysis with biological insights.} This novel method maps latent variables to original data, identifying key neurons with distinct temporal patterns and stimulus discriminative capabilities, while revealing visual regions sensitive to edges and luminance.
\end{itemize}

\section{Related Work}

\textbf{Latent Variable Modeling.} The advent of calcium imaging has provided unprecedented access to neural data while simultaneously introducing significant analytical challenges. A prominent approach to managing this complexity involves mapping high-dimensional neural signals to low-dimensional latent spaces. Early research focused on extracting latent variables from neural population activities using simple, prior-driven designs \cite{yu2008gaussian, zhao2017variational}. This evolved into more sophisticated methods employing recurrent neural network (RNN)-based explicit dynamics modeling \cite{pandarinath2018inferring, keshtkaran2022large, zhu2022deep}, which yielded highly interpretable results. The increasing scale of neural activity data led to the development of more efficient approaches, including Transformer-based methods with implicit latent representation learning \cite{ye2021representation, liu2022seeing, le2022stndt, ye2024neural, antoniades2023neuroformer}. While these demonstrated superior decoding performance, they often lacked interpretability and identifiability. To address these limitations, interpretable approaches such as pi-VAE \cite{zhou2020learning} and CEBRA \cite{schneider2023learnable} introduced auxiliary variables for identifiable latent variables, building upon identifiable variational autoencoders (iVAE) \cite{khemakhem2020variational} and nonlinear independent component analysis (ICA) \cite{hyvarinen2019nonlinear}. However, these methods prove inadequate for direct cross-individual analysis and fail to account for individual variability. Our approach differs by explicitly modeling cross-individual variability and separating neural activity components into stimulus-related and unrelated elements. Through multi-modal and bi-directed generative modeling, we effectively isolate highly relevant latent variables between neural activity and visual stimuli, ensuring interpretability while enabling sophisticated latent space modeling and analysis.

\begin{figure*}[t]
    \centering
    \includegraphics[scale=0.8]{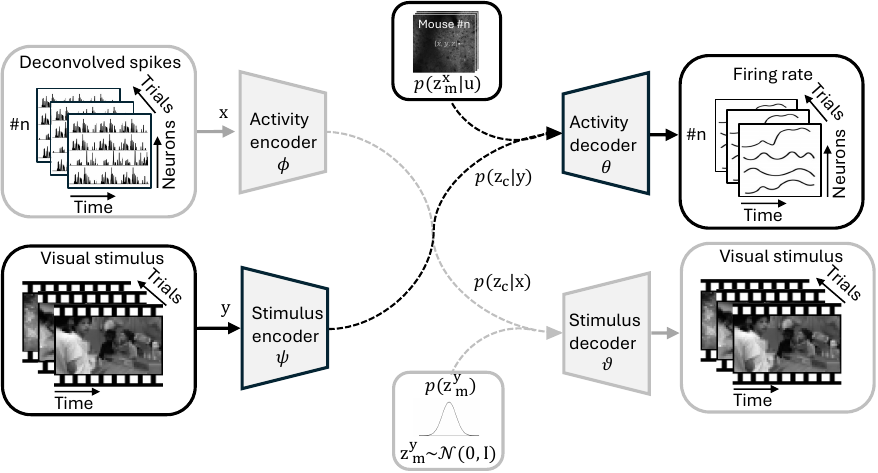}
    \caption{\textbf{miVAE for modeling V1 neural activity $\mathbf{x}$ and the corresponding visual stimulus $\mathbf{y}$.} $\mathbf{x}$ and $\mathbf{y}$ are decomposed in the latent space. For $\mathbf{x}$, it has the idiosyncratic latent variable $\mathbf{z_m^x}$ and the cross-subject and functionally relevant preserved latent variable $\mathbf{z_c^x}$. For $\mathbf{y}$, it has the neural activity-relevant variable $\mathbf{z_c^y}$ and the irrelevant variable $\mathbf{z_m^y}$. Our method employ generative modeling to derive these variables, with special focus on acquiring the most correlated variable $\mathbf{z_c^x}$, $\mathbf{z_c^y}$.}
    \label{mibivae-pipeline}
\end{figure*}

\textbf{Multi-Modal VAEs.} In machine learning, multi-modal VAEs aim to learn joint posterior approximations across modalities \cite{wu2018multimodal,shi2019variational,sutter2021generalized}. Traditional approaches often rely on restrictive assumptions, particularly regarding latent space aggregation \cite{sutter2021generalized}. Recent developments \cite{palumbo2023mmvae_p} have introduced modality-specific latent subspaces to enhance generative quality, while the latest MMVE \cite{sutter2024unity} implements data-dependent prior distributions for improved posterior approximation regulation. In neuroscience applications, emerging models like MM-GPVAE \cite{gondur2023multi} share our miVAE's latent space partition strategy. However, while these methods aggregate latent variables of neural activity and behavior, our approach implements directed generative modeling where neural activity and visual stimuli serve as mutual priors, avoiding fusion operations and emphasizing cross-individual modeling.

\textbf{Functional Modeling and Stimuli Reconstruction of V1.} Understanding V1 requires both functional modeling and neural activity decoding. Neural networks have successfully mapped visual stimuli to neural activity, revealing encoding mechanisms \cite{sinz2018stimulus, ecker2018rotation, bashiri2021flow, lurz2020generalization, ma2024temporal}. Wang et al. \cite{wang2023towards} proposed a shared feature extractor with separate readouts for V1 modeling, partially addressing the one-to-many challenge, though still requiring individual-specific fine-tuning. Current models often incorrectly assume complete visual information encoding in local V1 activity \cite{hubel1977ferrier, tootell1982deoxyglucose}. Decoding models focus on visual stimuli reconstruction from neural activity \cite{cobos2022takes, ellis2018high, yoshida2020natural}, revealing interpretable variance \cite{berens2012fast, eichhorn2003prediction, froudarakis2014population, garasto2018visual}. Recent algorithms employ both linear \cite{ellis2018high, yoshida2020natural} and nonlinear approaches \cite{cobos2022takes}, yet maintain the assumption of complete visual information encoding in local V1 activity. Our approach acknowledges that recorded local V1 activity contains only partial visual information \cite{olshausen2004sparse, roe2012toward} and utilizes relevant latent variables from multi-modal modeling for both encoding and decoding. Additionally, we introduce a novel score-based attribution strategy for mapping latent variables to original data, providing new insights into neural processing mechanisms.

\section{Method}
Our method comprises three components, specifically, \textbf{Cross-Individual Multi-Modal and Cross-Modal Modeling} to extract relevant hidden variables between neural activity and visual stimulation, \textbf{Neural Encoding and Decoding in Latent Space.} for further representation alignment, and \textbf{Score-based Attribution Analysis} for interpretative analysis.

\subsection{Cross-Individual Multi-Modal Modeling}
We present a framework for modeling multi-modal data consisting of neural activity $\mathbf{x} \in \mathbb{R}^{N \times T}$ and visual stimulus $\mathbf{y} \in \mathbb{R}^{H \times W \times T}$ across individuals. The framework addresses both modalities through complementary approaches. 

For neural activity $\mathbf{x}$, we derive bi-identifiable latent variables $\mathbf{z^x} \in \mathbb{R}^{d \times T}$ ($d \leq N$), comprising idiosyncratic latent variables $\mathbf{z_m^x} \in \mathbb{R}^{d/2 \times T}$ and cross-individual preserved latent variables $\mathbf{z_c^x} \in \mathbb{R}^{d/2 \times T}$. Individual-specific information $\mathbf{u}$ enhances the identifiability of $\mathbf{z_m^x}$, while visual stimulus $\mathbf{y}$ serves as a functional prior for $\mathbf{z_c^x}$. 

For visual stimulus $\mathbf{y}$, we construct semi-identifiable latent variables $\mathbf{z^y} \in \mathbb{R}^{d \times T}$, separated into activity-related latent variables $\mathbf{z_c^y}\in \mathbb{R}^{d/2 \times T}$ and activity-irrelevant latent variables $\mathbf{z_m^y} \in \mathbb{R}^{d/2 \times T}$. Thus, the latent variables $\mathbf{z_c^y}$ and $\mathbf{z_c^x}$ are highly correlated, which means there exists the ideal $\mathbf{z_c}$ derived either from $\mathbf{x}$ or $\mathbf{y}$.

The multi-modal generative latent modeling for $p(\mathbf{x,y|u})$ employs two complementary generative models, [1] \textbf{activity-based} and [2] \textbf{stimulus-based}:
\begin{equation}
    \begin{aligned}
        &p(\mathbf{x,y,z_m^x,z_c,z_m^y|u}) \\
        &= p_{\boldsymbol{\theta}}(\mathbf{x|z_m^x,z_c})  p_{\mathbf{T}, \boldsymbol{\lambda}}(\mathbf{z_m^x|u}) p_{\boldsymbol{\psi}}(\mathbf{z_c|y}) p(\mathbf{y,z_m^y}) \ [1]\\
        &=p_{\boldsymbol{\vartheta}}(\mathbf{y|z_c,z_m^y})  p_{\boldsymbol{\phi}}(\mathbf{z_c|x}) p(\mathbf{z_m^y}) p(\mathbf{x, z_m^x | u}) \ [2]
    \end{aligned}   
    \label{eq:prior}
\end{equation}
As shown in Figure \ref{mibivae-pipeline}, neural activity $\mathbf{x}$ and visual stimuli $\mathbf{y}$ serve as mutual priors, connected through $\mathbf{z_c}$.

In the \textbf{activity-based} model, we first define the prior for the idiosyncratic latent variable $\mathbf{z_m^x}$. Drawing from the universal approximation capabilities of exponential families \cite{sriperumbudur2017density}, we employ a factorized exponential family distribution conditioned on $\mathbf{u}$ \cite{khemakhem2020variational}:
\begin{equation}
    p_{\textbf{T},\boldsymbol{\lambda}}(\mathbf{z_m^x|u}) = \prod_{i=1}^m \frac{Q_i(\mathbf{z_m^x}_i)}{Z_i(\mathbf{u})}\exp[\sum_{j=1}^k T_{i,j}(\mathbf{z_m^x}_i)\lambda_{i,j}(\mathbf{u})]
    \label{eq:prior-ilv}
\end{equation}
where $Q_i$ denotes the base measure, $z_m^x(\mathbf{u})$ represents the normalizing constant, $\mathbf{T}_{i}=(T_{i,1}, \ldots, T_{i,k})$ comprises sufficient statistics, and $\boldsymbol{\lambda}_{i}(\mathbf{u})=(\lambda_{i,1}(\mathbf{u}), \ldots, \lambda_{i,k}(\mathbf{u}))$ represents the corresponding parameters. The dimension $k$ of each sufficient statistic is predefined, with $\boldsymbol{\lambda}(\cdot)$ parameterized through a neural network.

Then, the prior for the key latent variable $\mathbf{z_c^x}$ reflects the functional uniformity of V1 activity across individuals \cite{safaie2023preserved}. Given that recorded V1 activity is stimulus-evoked and functionally consistent across individuals, we employ a factorized exponential family distribution $p_{\boldsymbol{\psi}}(\mathbf{z_c|y})$, with statistics inferred through network $\boldsymbol{\psi}$.

For the \textbf{stimulus-based} modeling, we first introduce the prior of the activity-related latent variable, $\mathbf{z_c^y}$. Due to retinotopic mapping, only part of the visual stimulus information is related to the recorded local V1 activity \cite{hubel1977ferrier, tootell1982deoxyglucose}. Therefore, in this stimulus-based modeling, we aim to extract the latent variable most correlated between neural activity $\mathbf{x}$ and visual stimulus $\mathbf{y}$. We use $p_{\boldsymbol{\phi}}(\mathbf{z_c|x})$ as the prior, and parameterize it with network $\boldsymbol{\phi}$.

Furthermore, we account for the prior of the activity-irrelevant latent variable, $\mathbf{z_m^y}$. Given that receptive fields cannot be treated as specific, easily processable data (such as identity-related information) that neural networks can readily handle, we instead introduce a standard normal distribution prior for $p(\mathbf{z_m^y})$, specifically $\mathbf{z_m^y} \sim \mathcal{N}(\mathbf{0,1})$, as a substitute. Note that we emphasize that if more tractable prior information regarding receptive fields becomes available, the current modeling framework can be entirely restructured into a learnable, parameterized prior, analogous to the idiosyncratic latent variable $\mathbf{z_m^x}$.

\subsection{Directed Cross-Modal Generative Modeling}
The intrinsic correlation between the recorded neural activity $\mathbf{x}$ and visual stimulus $\mathbf{y}$ motivates our cross-modal modeling approach. Building on established neurophysiological principles \cite{hubel1962receptive, niell2008highly}, we model the stimulus-evoked V1 neural activity as:
\begin{equation}
    p(\mathbf{x|u,y}) = p_{\boldsymbol{\theta}}(\mathbf{x|z_m^x,z_c})  p_{\mathbf{T}, \boldsymbol{\lambda}}(\mathbf{z_m^x|u})p_{\boldsymbol{\psi}}(\mathbf{z_c|y})
    \label{ca2stimulus}
\end{equation}

Conversely, acknowledging that neural activity encodes visual stimulus information, we model this relationship in the latent space while accounting for activity-independent stimulus components:
\begin{equation}
    p(\mathbf{y|x}) = p_{\boldsymbol{\vartheta}}(\mathbf{y|z_c,z_m^y})  p_{\boldsymbol{\phi}}(\mathbf{z_c|x}) p(\mathbf{z_m^y})
    \label{stimulus2ca}
\end{equation}

This bidirectional modeling approach, expressed through Equations \ref{ca2stimulus} and \ref{stimulus2ca}, reinforces the correlation between neural activity and visual stimuli, particularly in extracting the shared components $\mathbf{z_c}$ in the latent space, as validated by experimental results.

\subsection{Tackling Dual Modeling via Variational Inference}
We adopt variational inference \cite{kingma2013auto} to approximate the intractable true posteriors $p(\mathbf{z_m^x|x, u})$, $p(\mathbf{z_c|x, y})$, and $p(\mathbf{z_m^y|y})$ with $q(\mathbf{z_m^x|x,u})$, $q(\mathbf{z_c|x,y})$, and $q(\mathbf{z_m^y|y})$ respectively. To optimize both the model and the approximate posteriors, we maximize the combined variational evidence lower bound (ELBO) of $p(\mathbf{x,y|u})$, $p(\mathbf{x|y,u})$, and $p(\mathbf{y|x})$, which is equivalent to minimizing the following:
\begin{equation}
    \begin{aligned}
        \mathcal{L}(\boldsymbol{\theta,\phi,\vartheta,\psi}) = \mathcal{L}_{\text{MM}}+\mathcal{L}_{\text{CM}} 
    \end{aligned}
\end{equation}
where $\mathcal{L}_{\text{MM}}$ represents the multi-modal loss and $\mathcal{L}_{\text{CM}}$ represents the cross-modal loss. Specifically, $\mathcal{L}_{\text{MM}}$ is as follows:
\begin{equation}
    \begin{aligned}
        &\mathcal{L}_{\text{MM}} =- \mathbb{E}_{q(\mathbf{z_m^x|x,u})q(\mathbf{z_c|x,y})}[\log p(\mathbf{x}| \mathbf{z_m^x,z_c})] \\
        &+ KL[q(\mathbf{z_m^x|x,u})||p(\mathbf{z_m^x|u})] + KL[q(\mathbf{z_c|x,y})||p(\mathbf{z_c|y})] \\
        &-\mathbb{E}_{q(\mathbf{z_c|x,y}) q(\mathbf{z_m^y|y})}[\log p(\mathbf{y|z_c,z_m^y})] \\
        &+ KL[q(\mathbf{z_m^y|y})||p(\mathbf{z_m^y})] + KL[q(\mathbf{z_c|x,y})||p(\mathbf{z_c|x})]
    \end{aligned}
\end{equation}
For the cross-modal loss $\mathcal{L}_{\text{CM}}$, which is responsible for enhancing correlation modeling, we have:
\begin{equation}
    \begin{aligned}
        &\mathcal{L}_{\textrm{CM}} = - \mathbb{E}_{p(\mathbf{z_m^x|u}) p(\mathbf{z_c|y})}[\log p(\mathbf{x|z_m^x,z_c})] \\
        &+ KL[p(\mathbf{z_m^x|u})||q(\mathbf{z_m^x|x,u})] + KL[p(\mathbf{z_c|y})||q(\mathbf{z_c|x,y})] \\
        &-\mathbb{E}_{p(\mathbf{z_c|x}) p(\mathbf{z_m^y})}[\log p(\mathbf{y|z_c,z_m^y})] \\
        &+ KL[p(\mathbf{z_c|x})||q(\mathbf{z_c|x,y})]
    \end{aligned}
\end{equation}
Detailed derivations are shown in Supplementary Materials. And we especially include the discussion of avoiding spillover effect for latent disentanglement in the Supplementary Materials. Next, we introduce the approximate posteriors for each latent variables.

\textbf{Approximate posterior of idiosyncratic latent variable $q(\mathbf{z_m^x|x,u})$.} 
For $\mathbf{z_m^x}$, we decompose the approximate posterior using a  factorized exponential family distribution, following \cite{johnson2016composing, zhou2020learning}
\begin{equation}
     q(\mathbf{z_m^x|x,u})\propto q_{\boldsymbol{\phi}}(\mathbf{z_m^x|x})p_{\mathbf{T},\boldsymbol{\lambda}}(\mathbf{z_m^x|u})
     \label{eq:post_ilv}
\end{equation}
where $q_{\boldsymbol{\phi}}(\mathbf{z_m^x|x})$ is assumed to be conditionally independent Gaussian distribution, i.e., $q_{\boldsymbol{\phi}}(\mathbf{z_m^x|x}) = \prod_{c=1}^m q(\mathbf{z_m^x}_{c}|\mathbf{x})$. Additionally, we assume that $q_{\boldsymbol{\phi}}(\mathbf{z_m^x|x})$ and $p_{\mathbf{T},\boldsymbol{\lambda}}(\mathbf{z_m^x|u})$ are independent distributions.

\begin{figure*}[t]
    \centering
    \includegraphics[scale=0.8]{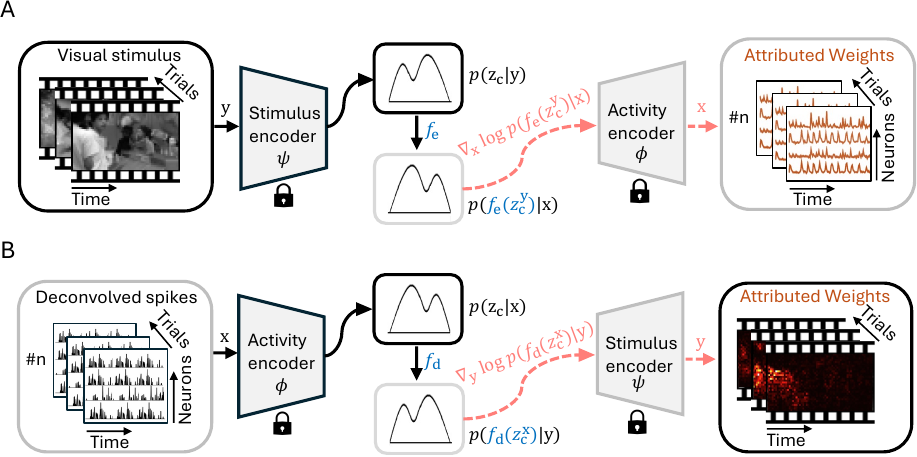}
    \caption{\textbf{Neural coding in the latent space with following score-based Attribution Analysis.} (A) The latent encoding function $f_{\text{e}}$ leverages pretrained miVAE encoders to align $\mathbf{z_c^x}$ and $\mathbf{z_c^y}$ in the shared latent space. Subsequently, the pretrained neural activity encoder $\boldsymbol{\phi}$ functions as a \textit{decoder}, enabling our score-based attribution analysis to identify salient neurons at each temporal point. (B) For latent decoding, an analogous approach using $f_{\text{d}}$ reveals key visual regions through the same attribution framework.}
    \label{fig:latent_encoding}
    \vspace{-1em}
\end{figure*}

\textbf{Approximate posterior of preserved latent variable $q(\mathbf{z_c|x, y})$.} Similar to $\mathbf{z_m^x}$, the approximate posterior for $\mathbf{z_c}$ is modeled using a factorized exponential family distribution. We assume conditionally independent Gaussian distributions $q_{\boldsymbol{\phi}}(\mathbf{z_c|x})$ and $p_{\boldsymbol{\psi}}(\mathbf{z_c|y})$.
\begin{equation}
    q(\mathbf{z_c|x,y}) \propto
    \begin{cases*}
        q_{\boldsymbol{\phi}}(\mathbf{z_c|x})p_{\boldsymbol{\psi}}(\mathbf{z_c|y}), & \ \ \text{for} \ \ $\mathbf{x}$ \\
        q_{\boldsymbol{\psi}}(\mathbf{z_c|y})p_{\boldsymbol{\phi}}(\mathbf{z_c|x}), & \ \ \text{for} \ \ $\mathbf{y}$
    \end{cases*}
    \label{eq:post_plv}
\end{equation}
It is important to note that for neural activity $\mathbf{x}$, the auxiliary variable is the visual stimulus $\mathbf{y}$, with $p_{\boldsymbol{\psi}}(\mathbf{z_c|y})$ as the prior. Conversely, for the visual stimulus $\mathbf{y}$, the auxiliary variable is the neural activity $\mathbf{x}$, with $p_{\boldsymbol{\phi}}(\mathbf{z_c|x})$ as the prior. Consequently, $q(\mathbf{z_c}|\mathbf{x}, \mathbf{y})$ serves as the shared approximate posterior for both $p(\mathbf{z_c}|\mathbf{y})$ and $p(\mathbf{z_c}|\mathbf{x})$. This consistency ensures that the latent variable $\mathbf{z_c}$ captures the shared information between neural activity $\mathbf{x}$ and visual stimulus $\mathbf{y}$. By aligning the latent variables produced by both encoders within a unified latent space, it facilitates the learning of robust and interpretable latent variables, which is crucial for subsequent encoding-decoding modeling and analysis.

\textbf{Approximate posterior of neural activity-irrelevant variable $q(\mathbf{z_m^y}|\mathbf{y})$.} The approximate posterior $q(\mathbf{z_m^y}|\mathbf{y})$ follows the same approach as the original VAE \cite{kingma2013auto}. It is a Gaussian distribution inferred by the visual stimulus encoding network $\boldsymbol{\phi}$ with input $\mathbf{y}$.

\subsection{Neural Encoding and Decoding in the Latent Space}
The above modeling enable us to acquire the highly related variable $\mathbf{z_c^x}$, $\mathbf{z_c^y}$ between the neural activity $\mathbf{x}$ and visual stimulus $\mathbf{y}$. Here, we further propose to modeling the encoding and decoding in the latent space, based on $\mathbf{z_c^x}$, $\mathbf{z_c^y}$. 

Given the pretrained miVAE, we have two encoders for neural activity $\mathbf{x}$ and visual stimulus $\mathbf{y}$ with parameters $\boldsymbol{\phi}$ and $\boldsymbol{\psi}$, respectively. Therefore, we have $\mathbf{z_c^x} \sim p_{\boldsymbol{\phi}}(\mathbf{z_c|x}) = \mathcal{N}(\boldsymbol{\mu}_{1}, \boldsymbol{\Sigma}_{1})$ and $\mathbf{z_c^y} \sim p_{\boldsymbol{\psi}}(\mathbf{z_c|y}) = \mathcal{N}(\boldsymbol{\mu}_{2}, \boldsymbol{\Sigma}_{2})$ \footnote{Note that the approximating posterior $q$ in miVAE are currently the known/prior distribution $p$ in latent coding modeling. Thus, for ease of illustration, we denote all distributions with $p$ in this latent coding section and following attribution analysis section.}. Here, we transform the original neural encoding and decoding into a distribution matching problem in the latent space. For neural encoding, we map $\mathbf{z_c^y} \sim p_{\boldsymbol{\psi}}(\mathbf{z_c|y})$ to $\mathbf{z_c^x} \sim p_{\boldsymbol{\phi}}(\mathbf{z_c|x})$, and similarly for neural decoding, as follows:
\begin{equation}
    \begin{cases*}
        \mathbf{z_c^x} = \mathbf{A}_{1}^T\mathbf{z_c^y} + \mathbf{b}_{1}, \mathbf{A}_{1}^{*}=\boldsymbol{\Sigma}_{1}^{1/2}\boldsymbol{\Sigma}_{2}^{-1/2}, \mathbf{b}_{1}^{*}= \boldsymbol{\mu}_{1}-\mathbf{A}_{1}^{*}\boldsymbol{\mu}_{2}, \\
        \mathbf{z_c^y} = \mathbf{A}_{2}^T\mathbf{z_c^x} + \mathbf{b}_{2}, \mathbf{A}_{2}^{*}=\boldsymbol{\Sigma}_{2}^{1/2}\boldsymbol{\Sigma}_{1}^{-1/2}, \mathbf{b}_{2}^{*}= \boldsymbol{\mu}_{2}-\mathbf{A}_{2}^{*}\boldsymbol{\mu}_{1}, \\
    \end{cases*}
\end{equation}
where $\mathbf{A}_{1}^{*}$, $\mathbf{b}_{1}^{*}$, $\mathbf{A}_{2}^{*}$ and $\mathbf{b}_{2}^{*}$ represent the optimal transform solutions for single paired ($\mathbf{z_c^x}$, $\mathbf{z_c^y}$). For cross-individual validation, these parameters are learned using either linear or nonlinear functions. We shown such encoding and decoding functions in the middle of Figure \ref{fig:latent_encoding}. During training, we use the KL-divergence that matching the transformed distribution and target distribution as the loss function. Modeling in latent space with unified latent size enable us to easily extend the cross-individual modeling and following analysis.

\subsection{Score-based Attribution Analysis}
Encoding and decoding in the latent space can further improve the correlations across $\mathbf{z_c^x}$ and $\mathbf{z_c^y}$. However, this does not yet provide a direct interpretation of the original data, which is critical for pellucid analysis. To solve this problem, we propose an new algorithm to trace back to the original data, based on the better-aligned latent variables, as shown in Figure \ref{fig:latent_encoding}.

Specifically, we present an attribution strategy based on a score function, treating the neural activity encoder as a \textit{decoder}. For $\mathbf{z_c^y}\rightarrow\mathbf{z_c^x}$, this attribution reveals the preferences of neuron subpopulation within a neural population for specific visual stimuli or measures the importance of these subpopulation for given visual stimuli at any given timestep. Similarly, for $\mathbf{z_c^x}\rightarrow\mathbf{z_c^y}$, we treat the visual stimulus encoder as a \textit{decoder} to analyze which parts of the visual stimulus are critical for the given neural activity. 

To illustrate our attribution strategy, we illustrate the principles behind the latent encoding $\mathbf{z_c^y}\rightarrow\mathbf{z_c^x}$ attribution strategy. We have the transformed distribution for $\mathbf{x}$ given by $p(f_{\text{e}}(\mathbf{z_c^y)|y})\approx p(f_{\text{e}}(\mathbf{z_c^y)|x})$. By applying Baysian rule, $p(f_{\text{e}}(\mathbf{z_c^y})|\mathbf{x}) = \frac{p(\mathbf{x}|f_{\text{e}}(\mathbf{z_c^y}))p(f_{\text{e}}(\mathbf{z_c^y}))}{p(\mathbf{x})}$, then we have following score function:
\begin{equation}
    \begin{aligned}
        \nabla_{\mathbf{x}} \log p(f_{\text{e}}&(\mathbf{z_c^y})|\mathbf{y})  \approx \nabla_{\mathbf{x}} \log p(f_{\text{e}}(\mathbf{z_c^y})|\mathbf{x}) \\
        &=\nabla_{\mathbf{x}} \log p(\mathbf{x}|f_{\text{e}}(\mathbf{z_c^y})) - \nabla_{\mathbf{x}} \log p(\mathbf{x})
    \end{aligned}
\end{equation}
where $\nabla_{\mathbf{x}} \log p(f_{\text{e}}(\mathbf{z_c^y})|\mathbf{x})$ represents the score \cite{fisher1970statistical} of the transformed latent distribution $f_{\text{e}}(\mathbf{z_c^y})$ comprises two key components. The first term, $\nabla_{\mathbf{x}} \log p(\mathbf{x}|f_{\text{e}}(\mathbf{z_c^y}))$, represents the gradient of the log-likelihood with respect to observed neural activity $\mathbf{x}$, given the transformed latent variable (distribution). The second term, $\nabla_{\mathbf{x}} \log p(\mathbf{x})$, acts as a correction factor through the marginal log-likelihood gradient. This formulation enables $\nabla_{\mathbf{x}} \log p(f_{\text{e}}\mathbf{(z_c^y) | x})$ to quantify neuronal importance relative to $\mathbf{z_c^y}$ at any timestep. The same principles extend to latent decoding ($\mathbf{z_c^x}\rightarrow\mathbf{z_c^y}$), providing a comprehensive framework for attribution analysis in both encoding and decoding processes.

In summary, the latent variables 
$\mathbf{z_c^x}$, $\mathbf{z_c^y}$ and learned $f_{\text{e}}$ and $f_{\text{d}}$ serve as bridges connecting neural activity $\mathbf{x}$ and visual stimuli $\mathbf{y}$. The proposed score function analysis enable us get rid of the prior of true distributions of $\mathbf{z_c^x}$ and $\mathbf{z_c^y}$, and can generate new insights into the original data space, potentially serving as a new alternative analysis tool for neural activity.

\section{Experiments}
We evaluated our approach using single-trial calcium imaging data from the Sensorium 2023 dataset \cite{turishcheva2023dynamic}, comprising neural recordings from 10 head-fixed mice paired with corresponding visual stimuli. The dataset included 5 distinct mouse pairs, where each pair viewed identical visual sequences. Neural activity was recorded via two-photon calcium imaging \cite{sofroniew2016large} and subsequently deconvolved into spike trains \cite{turishcheva2023dynamic}, yielding recordings from 78,853 neurons. All recordings were temporally aligned at 30Hz sampling rate, with each trial spanning approximately 10 seconds. The visual stimuli consisted of movies at $36\times64$ pixel resolution.

Data from 7 mice were used for training, with the remaining 3 mice reserved for validation. All results are reported on previously unseen mice. Detailed experimental protocols are provided in the Supplementary Materials.

\begin{table}[t]
\centering
\setlength{\tabcolsep}{0.45mm}{
    \begin{tabular}{lccccc}
    \toprule
     & &\multicolumn{4}{c}{\textbf{Stage 2}} \\
    \cmidrule(lr){3-6}
    \textbf{Methods} &\textbf{Stage 1} & \multicolumn{2}{c}{\textbf{Latent Encoding}} & \multicolumn{2}{c}{\textbf{Latent Decoding}} \\
    \cmidrule(lr){3-4} \cmidrule(lr){5-6}
    & & \textbf{Linear} & \textbf{nonLinear} & \textbf{Linear} & \textbf{nonLinear} \\
    \hline
    \rowcolor{gray!5}
    V1-FM$_{\mathbf{e}}^{*}$ &0.2169* & - &-  & - &-\\
    \rowcolor{gray!5}
    V1-FM$_{\mathbf{d}}^{*}$ &0.3626* & - & - & - &- \\
    \hline
    VAE$\dag$ & 0.0004 & 0.1869 & 0.1989 & 0.0074 & 0.0331 \\
    iVAE$\dag$ & 0.1677 & 0.6416 & 0.8343 & 0.4852 & 0.8341 \\
    MVAE & 0.7992  &0.8023  &0.8393  &0.7983  &0.8641  \\
    MMVAE & 0.0067 & 0.0048 &0.0339 &0.0294 & 0.0930 \\
    MoPoE & -0.0093& 0.0123 & 0.0563 & 0.0257 &  0.1215\\ 
    MEME & 0.7746  &0.7821 & 0.7963 & 0.7785 & 0.8692 \\
    MMVE & 0.7160  & 0.7306 & 0.7538 & 0.7206 & 0.8304 \\ 
    \textbf{miVAE} &\textbf{0.8694} &\textbf{0.8809} &\textbf{0.9149} & \textbf{0.8770} &\textbf{0.9094}\\
    \bottomrule
    \end{tabular}
}
\caption{\textbf{Comparisons of neural encoding and decoding.} We report correlation on the latent variables $\mathbf{z_c^x}$ and $\mathbf{z_c^y}$ generated by the two encoders in VAE models and further coding models. * indicates the coding models are in the original data space. $\dag$ indicates the VAE in Stage 1 are separately trained.}
\label{tab:main_experiment}
\vspace{-1em}
\end{table}

\subsection{Baseline and Evaluation Metric}
\textbf{Baseline.} We considered two types of baseline models. The first type codes in the original data space, using a core and linear readout module \cite{lurz2020generalization}. Following \cite{wang2023towards}, we pre-trained the encoding model on data from seven mice, then fixed the core and fine-tuned the readout on data from three mice, referred to as V1-FM$_{\text{e}}$. The same method was used for decoding, referred to as V1-FM$_{\text{d}}$.

The second type codes in the latent space using a two-stage process. First, we applied self-supervised models to neural activity or visual stimuli using VAE \cite{kingma2013auto} and iVAE \cite{zhou2020learning}. We also consider multi-modal VAEs in machine learning, including MVAE \cite{wu2018multimodal}, MMVAE \cite{shi2019variational}, MoPoE \cite{sutter2021generalized}, MEME \cite{joy2021learning} and MMVE \cite{sutter2024unity}. Second, we performed linear or nonlinear coding in the latent space with $d=16, T=256$. All models share identical architectures.

\textbf{Metric.} Following Wang et al. \cite{wang2023towards}, we employed the single-trial Pearson Correlation Coefficient (R) as our evaluation metric. For the reference V1-FM models, we measure the similarity between predicted and real neural activity for the encoding model, and between decoded and actual visual stimuli for the decoding model. In our latent space encoding-decoding framework, we first compare latent variables $\mathbf{z_c}$ from the visual stimulus encoder and the neural activity encoder for VAE models. In the second step, involving latent coding models, we fix the latent variable from one encoder as the reference and calculate the correlation of the coding model's output with this reference. We use this setting for all quantitative results and ablation studies.

\subsection{Quantitative Results for Neural Coding.}
\textbf{miVAE and neural latent modeling achieve remarkable coding performance.} Table \ref{tab:main_experiment} shows the performance of encoding and decoding models. Though coding in the original data space cannot be directly compared to latent space coding, our approach demonstrates more effective coding than similar baselines. miVAE and subsequent latent coding surpasses all baseline models, highlighting the effectiveness of our whole method. The validation was performed on $\mathbf{z_c}$ of $\mathbf{x}$ and $\mathbf{y}$.

\textbf{miVAE and latent modeling shows outstanding cross-individual alignment.} Table \ref{tab:model_consistency} shows correlations between pairs of mice (5 pairs) under the same visual stimuli. Previous multi-modal VAEs achieve partial alignment. In contrast, our miVAE achieves state-of-the-art alignment performance in the first stage compared to former all results, and following decoding alignment further enhanced the performance, indicating high similarity in neural activity latent variables under the same stimuli \cite{safaie2023preserved}.

\begin{table}[t]
\centering
\setlength{\tabcolsep}{0.45mm}{
    \begin{tabular}{lccc}
    \toprule
     \multirow{2}{*}{\textbf{Methods}} & \multirow{2}{*}{\textbf{Stage 1}} & \multicolumn{2}{c}{\textbf{Latent Decoding (Stage 2)}}\\
     \cmidrule(lr){3-4}
     & & \textbf{Linear} & \textbf{NonLinear} \\
    \midrule
    VAE$\dag$ &0.0822$\pm$0.0094 &0.0371$\pm$0.0110 &0.1490$\pm$0.0285\\
    iVAE$\dag$ &0.7449$\pm$0.0111 &0.3345$\pm$0.0241 &0.7855$\pm$0.0098\\
    MVAE &0.7316$\pm$0.0091 &0.7286$\pm$0.0095 &0.7908$\pm$0.0027\\
    MMVAE & 0.1417$\pm$0.0242&0.1502$\pm$0.0227 &0.1902$\pm$0.0654  \\
    MoPoE & 0.1708$\pm$0.0312&0.1738$\pm$0.0366 &0.2043$\pm$0.0366  \\    
    MEME & 0.6505$\pm$0.0100&0.6526$\pm$0.0088 &0.7398$\pm$0.0086  \\
    MMVE & 0.6598$\pm$0.0038 &0.6651$\pm$0.0037 &0.7594$\pm$0.0027 \\    
    \textbf{miVAE}  & \textbf{0.8984}$\pm$0.0159 & \textbf{0.9091}$\pm$0.0133 & \textbf{0.9635}$\pm$0.0067 \\
    \bottomrule
    \end{tabular}
}
\caption{\textbf{Correlation between paired mice with the same visual stimuli.} We report the mean and standard deviation of the correlation of $\mathbf{z_c^x}$ across all 5 pairs of mice.}
\label{tab:model_consistency}
\end{table}

\begin{figure}[ht]
    \centering
    \includegraphics[scale=0.485]{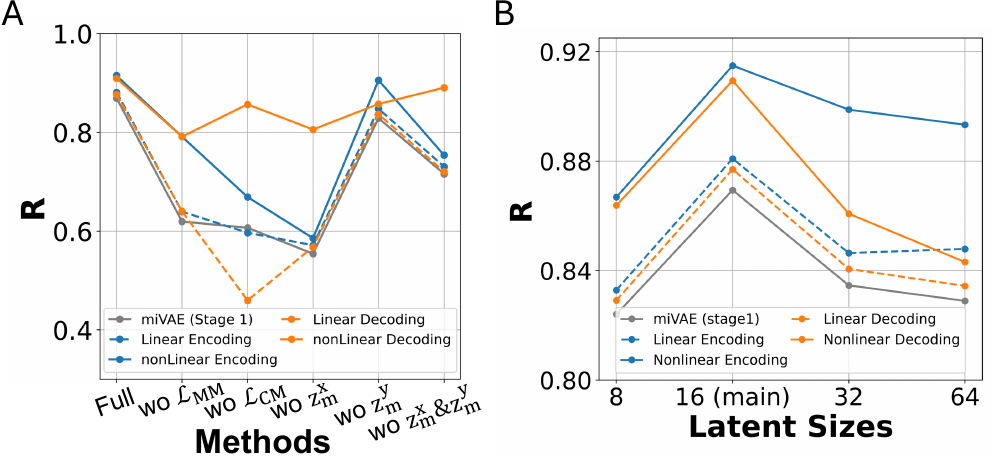}
    \caption{\textbf{Ablations on modeling methods.} (A) Ablations on losses and latent variables. (B) Ablations on Latent size. *R refers to the correlation of $\mathbf{z_c^x}$ and $\mathbf{z_c^y}$.}
    \label{fig:ablations_model}
    \vspace{-2.1mm}
\end{figure}

\section{Ablation studies} 
\textbf{Ablations on modeling methods.} In Figure \ref{fig:ablations_model}.A, we show that combining $\mathcal{L}_{\textrm{CM}}$ with $\mathcal{L}_{\textrm{MM}}$ yields the best performance. Using only $\mathcal{L}_{\textrm{CM}}$ generally surpasses using only $\mathcal{L}_{\textrm{MM}}$, indicating better regularization on latent variables.

Ablation studies on latent variables demonstrate that omitting $\mathbf{z_m^x}$ while retaining $\mathbf{z_m^y}$, or excluding both, substantially impairs performance, highlighting the critical role of cross-individual modeling. Notably, retaining $\mathbf{z_m^x}$ while omitting $\mathbf{z_m^y}$ maintains robust performance, highlighting the exceptional efficacy of $\mathbf{z_m^x}$ in capturing cross-individual variations. We suppose that the minimal impact of $\mathbf{z_m^y}$ stems from no priors for its modeling.

In Figure \ref{fig:ablations_model}.B, we demonstrate that a latent size of 16 is optimal, with the best generalization error, and used in our main method.

\begin{figure}[t]
    \centering
    \includegraphics[scale=0.32]{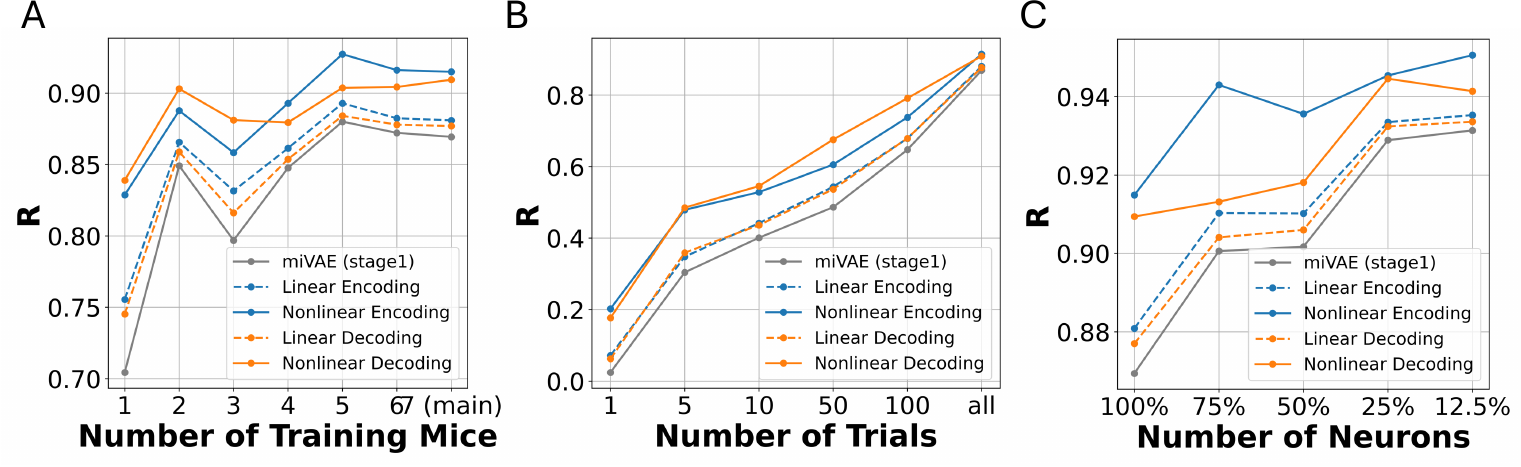}
    \caption{\textbf{Ablations studies on data scale.} (A) R over increasing training mice number. (B) R over increasing training trials. (C) R over different number of neurons, decreasing from L2/3 to L5. *R refers to the correlation of $\mathbf{z_c^x}$ and $\mathbf{z_c^y}$.}
    \label{fig:ablations_data}
\end{figure}

\textbf{Ablations on data scale.} We further considered the impact of the training data scale. As illustrated in Figure \ref{fig:ablations_data}.A, increasing the number of mice in the training set correlates positively with performance. Similarly, Figure \ref{fig:ablations_data}.B shows that increasing the number of training trials yields improved correlations. These findings indicate the beneficial effect of increased data volume on cross-individual generalization.

More importantly, Figure \ref{fig:ablations_data}.C reveals an relationship between neuronal population size and model performance. Smaller neuron counts in deeper layers, e.g., L5, yield higher correlations, showing our model's ability in modeling small-scale neural activity. Besides, this results also support sparse neural coding principles in visual systems \cite{yoshida2020natural}, and suggest deeper layer neurons might be particularly effective for stimulus encoding. 

\section{Attribution Analysis based on Latent Coding}
\begin{figure}[t]
    \centering
    \includegraphics[scale=0.56]{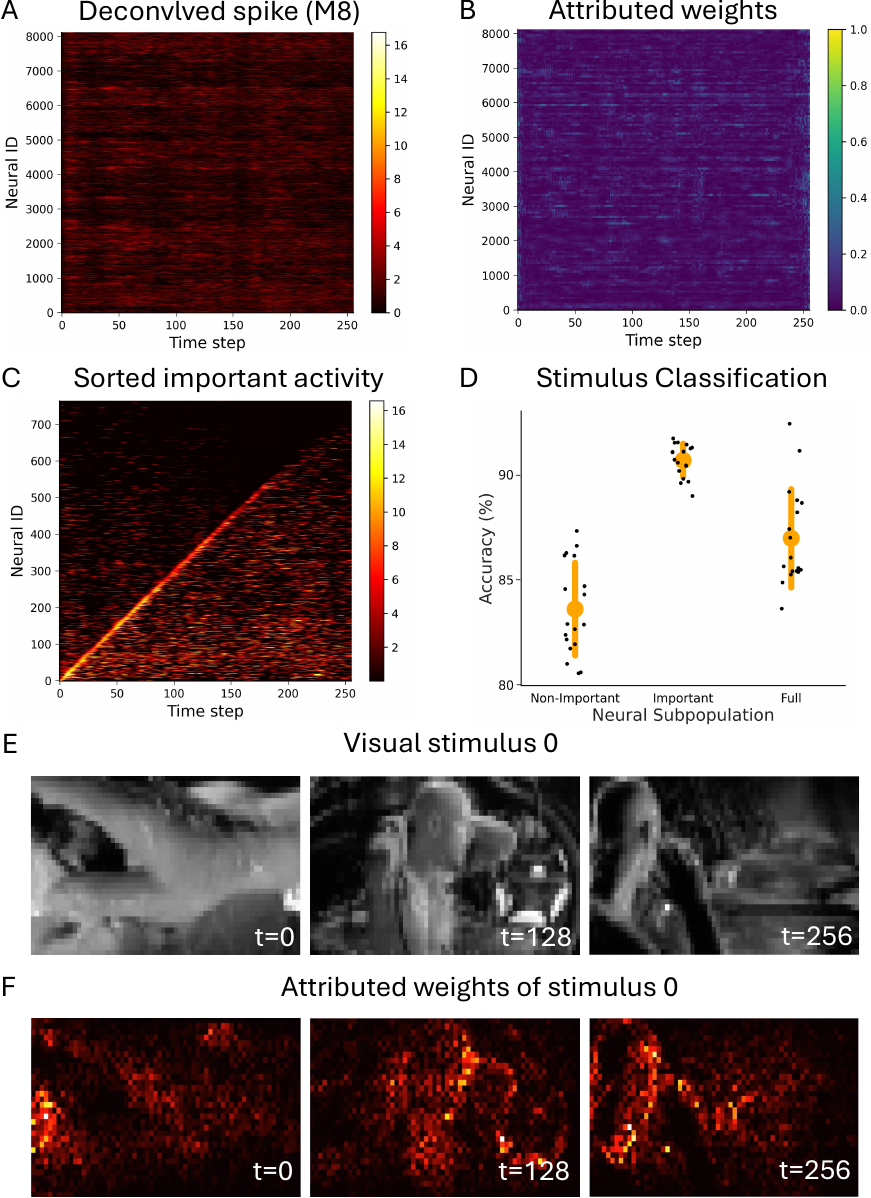}
    \caption{\textbf{Bidirectional Attribution analysis.} (A) Recorded neural activity of Mouse 8 (M8), $\mathbf{x}$. (B) Attributed weights of $\mathbf{x}$ based on $\mathbf{z_c^y}$. (C) Sorted important activity shows noteworthy temporal pattern. (D) Stimulus classification shows the identified key subpopulations best at distinguishing visual stimuli. (E) Visual stimulus samples, $\mathbf{y}$. (F) Attributed weights of $\mathbf{y}$ based on $\mathbf{z_c^x}$ shows the recorded activity has high correlation with edge and luminance.}
    \label{fig:attribution_analysis}
\end{figure}

We conducted score-based attribution analysis using the high-quality latent variable $\mathbf{z_c}$ with encoding or decoding functions $f_{\text{e}}$,$f_{\text{d}}$. For the neural activity of Mouse 8 in the validation set (Figure \ref{fig:attribution_analysis}.A), we attributed the encoded visual stimuli $\mathbf{z_c^y}$, obtaining the neuron weights shown in Figure \ref{fig:attribution_analysis}.B, which range from 0 to 1 at any given time. We then identified important neurons as those with weights greater than 0.55 at all timesteps. By sorting the activities of these important neurons, we observed the significant pattern shown in Figure \ref{fig:attribution_analysis}.C, indicating the temporal patterns of important neurons responsive to visual stimuli. 

We further validated the classification of visual stimuli using the identified important neurons from three mice in the validation set. We use 7 classical classifiers, including Linear SVM, RBF SVM, Polynomial SVM, KNN (with n$\_$neighbors=2), Decision Tree, Random Forest, Naïve Bayes, and MLP, and report their averaged performance. Figure \ref{fig:attribution_analysis}.D shows that the neural activity of these important neuron sub-populations (91.29$\%$ with about 700 neurons) significantly outperformed that of the non-important neurons (82.92$\%$ with more than 7000 neurons) and even exceeded the results of the full population (87.24$\%$). This confirms the significance of the identified important neurons. Detailed implementations are provided in Supplementary Materials.

Lastly, we performed attribution analysis of the decoded latent variable $\mathbf{z_c^x}$ to the visual stimuli (Figure \ref{fig:attribution_analysis}.E). This analysis revealed that the neural activity exhibited a stronger correlation with edge and luminance (Figure \ref{fig:attribution_analysis}.F).

\section{Conclusions}
We present a multi-modal identifiable VAE, miVAE, and latent modeling, acquiring the highly correlated variable between neural activity and visual stimulus. By further combining the proposed score-based attribution analysis, we reverse to the original data space to identify important neurons and visual regions. The whole method shows promising across-individual coding, alignment and fine-grained analysis for V1. With high precision and interpretability, our approach holds significant potential for modeling and analyzing neural data from complex sensory cortices, such as the auditory cortex.


\bibliography{aaai25}
\newpage
\section{ELBO derivation for Multi-Modal Modeling}
\label{SI:elbo-mm}
\subsection{Notations}
For neural activity $\mathbf{x}$, we define its latent variables $\mathbf{z^x} = [\mathbf{z_m^x}, \mathbf{z_c^x}]$, where $\mathbf{z} \in \mathbb{R}^d$, $\mathbf{z_m^x} \in \mathbb{R}^{d/2}$, and $\mathbf{z_c} \in \mathbb{R}^{d/2}$. The variables $\mathbf{z_m^x}$ and $\mathbf{z_c^x}$ are independent due to their different prior information, which comes from significantly distinct domain.

For visual stimulus $\mathbf{y}$, we define the latent variables $\mathbf{z^y} = [\mathbf{z_c^y}, \mathbf{z_m^y}]$, where $\mathbf{z'} \in \mathbb{R}^d$, $\mathbf{z_c^y} \in \mathbb{R}^{d/2}$, and $\mathbf{z_m^y} \in \mathbb{R}^{d/2}$. The variables $\mathbf{z_c^y}$ and $\mathbf{z_m^y}$ are also assumed to be independent, as we aim to identify the variables related and unrelated to the visual stimulus.

Note that the ideal $\mathbf{z_c}$ for both $\mathbf{x}$ and $\mathbf{y}$ shares the same prior distribution according our assumption.

\subsection{Activity-based ELBO}
In this section, we derive the Evidence Lower Bound (ELBO) based on neural activtiy modeling, as follows:
\begin{equation}
\resizebox{\linewidth}{!}{$
    \begin{aligned}
        &\log \ p(\mathbf{x,y|u}) = \log \int\int p(\mathbf{x,y,z_m^x,z_c|u}) d\mathbf{z_m^x}d\mathbf{z_c}\\
        &= \log \int\int p(\mathbf{x|z_m^x,z_c})  p(\mathbf{z_m^x|u}) p(\mathbf{z_c|y}) p(\mathbf{y, z_m^y}) d\mathbf{z_m^x}d\mathbf{z_c}\\
        &= \log \int\int \frac{p(\mathbf{x|z_m^x,z_c}) p(\mathbf{z_c|y}) p(\mathbf{z_m^x|u}) p(\mathbf{y, z_m^y}) }{q(\mathbf{z_m^x|x,u})q(\mathbf{z_c|x,y})}q(\mathbf{z_m^x|x,u}) q(\mathbf{z_c|x,y}) d\mathbf{z_m^x}d\mathbf{z_c} \\
        &\geq \mathbb{E}_{q(\mathbf{z_m^x|x,u}) q(\mathbf{z_c|x,y})}[\log \frac{p(\mathbf{x|z_m^x,z_c}) p(\mathbf{z_m^x|u}) p(\mathbf{z_c|y}) p(\mathbf{y, z_m^y}) }{q(\mathbf{z_m^x|x,u}) q(\mathbf{z_c|x,y})}]\\
        &= \mathbb{E}_{q(\mathbf{z_m^x|x,u})q(\mathbf{z_c|x,y})}[\log p(\mathbf{x}| \mathbf{z_m^x,z_c})] + \mathbb{E}_{q(\mathbf{z_m^x|x,u})}[\frac{p(\mathbf{z_m^x|u})}{q(\mathbf{z_m^x|x,u})}] \\
        &+  \mathbb{E}_{q(\mathbf{z_c|x,y})}[\frac{p(\mathbf{z_c|y})}{q(\mathbf{z_c|x,y})}] + \log p(\mathbf{y, z_m^y})\\
        &=\mathbb{E}_{q(\mathbf{z_m^x|x,u})q(\mathbf{z_c|x,y})}[\log p(\mathbf{x}| \mathbf{z_m^x,z_c})] - KL[q(\mathbf{z_m^x|x,u})||p(\mathbf{z_m^x|u})]\\
        &-KL[q(\mathbf{z_c|x,y})||p(\mathbf{z_c|y})] + \log p(\mathbf{y, z_m^y})
    \end{aligned}
    $}
    \label{SI:elbo-activity}
\end{equation}

\subsection{Stimulus-based ELBO}
In this section, we derive the Evidence Lower Bound (ELBO) based on stimulus-based modeling, as follows:
\begin{equation}
\resizebox{\linewidth}{!}{$
    \begin{aligned}
        & \log p(\mathbf{x,y|u}) = \log \int\int p(\mathbf{x,y,z_m^y,z_c|u}) d\mathbf{z_m^y}d\mathbf{z_c} \\
        &= \log \int\int p(\mathbf{y|z_c,z_m^y})  p(\mathbf{z_c|x}) p(\mathbf{z_m^y}) p(\mathbf{x, z_m^x | u}) d\mathbf{z_m^y}d\mathbf{z_c}\\
        &= \log \int\int \frac{p(\mathbf{y|z_c,z_m^y})  p(\mathbf{z_c|x}) p(\mathbf{z_m^y}) p(\mathbf{x, z_m^x | u})}{q(\mathbf{z_c|x,y})q(\mathbf{z_m^y|y})}q(\mathbf{z_c|x,y})q(\mathbf{z_m^y|y}) d\mathbf{z_m^y}d\mathbf{z_c} \\
        &\geq \mathbb{E}_{q(\mathbf{z_c|x,y}) q(\mathbf{z_m^y|y})}[\log \frac{p(\mathbf{y|z_c,z_m^y})  p(\mathbf{z_c|x}) p(\mathbf{z_m^y}) p(\mathbf{x, z_m^x | u})}{q(\mathbf{z_c|x,y})q(\mathbf{z_m^y|y})}] \\
        &= \mathbb{E}_{q(\mathbf{z_c|x,y}) q(\mathbf{z_m^y|y})}[\log p(\mathbf{y|z_c,z_m^y})] - KL[q(\mathbf{z_c|x,y})||p(\mathbf{z_c|x})] \\
        &- KL[q(\mathbf{z_m^y|y})||p(\mathbf{z_m^y})] + \log p(\mathbf{x,z_m^x|u})
    \end{aligned}
    $}
    \label{SI:elbo-stimulus}
\end{equation}
where $p(\mathbf{z_m^y})$ is practically a factorized, isotropic unit Gaussian.

\subsection{Multi-modal Training Loss}
We have observed that Eq. \ref{SI:elbo-activity} and Eq. \ref{SI:elbo-stimulus} contain two intractable log-likelihood terms, for which precise analytical solutions are unattainable. Fortunately, when we combine the two ELBOs, the regularizing effects of these log-likelihood terms are effectively complemented by the ELBOs of the respective other modality. Therefore, we empirically employ the following multi-modal loss function for approximate optimization:
\begin{equation}
    \begin{aligned}
        &\mathcal{L}_{\textrm{MM}} = - \mathbb{E}_{q(\mathbf{z_m^x|x,u})q(\mathbf{z_c|x,y})}[\log p(\mathbf{x}| \mathbf{z_m^x,z_c})] \\
        &+ KL[q(\mathbf{z_m^x|x,u})||p(\mathbf{z_m^x|u})] + KL[q(\mathbf{z_c|x,y})||p(\mathbf{z_c|y})] \\
        &-\mathbb{E}_{q(\mathbf{z_c|x,y}) q(\mathbf{z_m^y|y})}[\log p(\mathbf{y|z_c,z_m^y})] \\
        &+ KL[q(\mathbf{z_m^y|y})||p(\mathbf{z_m^y})] + KL[q(\mathbf{z_c|x,y})||p(\mathbf{z_c|x})]
    \end{aligned}
    \label{SI:loss-mm}
\end{equation}

\section{ELBO derivation for Cross-Modal Modeling}
In addition to multi-modal modeling, we also consider the causal relationship where neural activity $\mathbf{x}$ is induced by visual stimulus $\mathbf{y}$. Similarly, there is information in $\mathbf{y}$ that is directly related to $\mathbf{x}$. To reinforce this causal relationship, we further explore cross-modal modeling.

We first consider the mapping from visual stimulus $\mathbf{y}$ to neural activity $\mathbf{x}$.
\begin{equation}
\resizebox{\linewidth}{!}{$
    \begin{aligned}
	\log p(\mathbf{x|u,y}) &= \log \int\int p(\mathbf{x,z_m^x,z_c|y,u}) d\mathbf{z_m^x}d\mathbf{z_c}\\
	&= \log \int\int p(\mathbf{x|z_m^x,z_c})  p(\mathbf{z_m^x|u})p(\mathbf{z_c|y}) d\mathbf{z_m^x}d\mathbf{z_c}\\
	&= \log \int\int \frac{p(\mathbf{x|z_m^x,z_c}) p(\mathbf{z_m^x|u}) p(\mathbf{z_c|y}) }{q(\mathbf{z_m^x|u})q(\mathbf{z_c|y})}q(\mathbf{z_m^x|u})q(\mathbf{z_c|y})  d\mathbf{z_m^x}d\mathbf{z_c}\\
	&\geq \mathbb{E}_{q(\mathbf{z_m^x|u}) q(\mathbf{z_c|y})}[\log \frac{p(\mathbf{x|z_m^x,z_c}) p(\mathbf{z_m^x|u}) p(\mathbf{z_c|y}) }{q(\mathbf{z_m^x|u}) q(\mathbf{z_c|y})}]\\
	&= \mathbb{E}_{q(\mathbf{z_m^x|u}) q(\mathbf{z_c|y})}[\log \frac{p(\mathbf{x|z_m^x,z_c}) p(\mathbf{z_m^x|u})p(\mathbf{z_c|y}) }{q(\mathbf{z_m^x|u}) q(\mathbf{z_c|y})}]\\
	&= \mathbb{E}_{q(\mathbf{z_m^x|u}) q(\mathbf{z_c|y})}[\log p(\mathbf{x|z_m^x,z_c})] \\
        &- KL[q(\mathbf{z_m^x|u})||p(\mathbf{z_m^x|u})] - KL[q(\mathbf{z_c|y})||p(\mathbf{z_c|y})] 
    \end{aligned}
    \label{eq:cm_1}
$}
\end{equation}
Given the paired nature of our data and building upon the previously trained multi-modal ELBO, we approximate the intractable prior $p(\mathbf{z_m^x|u})$ with $q(\mathbf{z_m^x|u,x})$. Similarly, we use $q(\mathbf{z_c|x,y})$ to approximate the prior distribution $p(\mathbf{z_c|y})$.

Symmetrically, we next examine the mapping from neural activity $\mathbf{x}$ to visual stimulus $\mathbf{y}$.
\begin{equation}
\resizebox{\linewidth}{!}{$
    \begin{aligned}
	\log p(\mathbf{y|x}) &= \log \int\int p(\mathbf{y,z_m^y,z_c|x})  d\mathbf{z_m^y}d\mathbf{z_c}\\
	&= \log \int\int p(\mathbf{y|z_c,z_m^y})  p(\mathbf{z_c|x}) p(\mathbf{z_m^y}) d\mathbf{z_m^y}d\mathbf{z_c}\\
    &= \log \int\int \frac{p(\mathbf{y|z_c,z_m^y})  p(\mathbf{z_c|x}) p(\mathbf{z_m^y})}{q(\mathbf{z_c|x})} q(\mathbf{z_c|x}) d\mathbf{z_m^y}d\mathbf{z_c}\\
	&\geq \mathbb{E}_{q(\mathbf{z_c|x})p(\mathbf{z_m^y})}[\log \frac{p(\mathbf{y|z_c,z_m^y})  p(\mathbf{z_c|x}) p(\mathbf{z_m^y})}{q(\mathbf{z_c|x})}]\\
    &= \mathbb{E}_{q(\mathbf{z_c|x}) p(\mathbf{z_m^y})}[\log p(\mathbf{y|z_c,z_m^y})] \\
    &- KL[q(\mathbf{z_c|x})||p(\mathbf{z_c|x})] + (-\frac{1}{2}\log 2\pi - \frac{1}{2})
    \end{aligned}
    \label{eq:cm_2}
$}
\end{equation}
Given the paired nature of our data and leveraging the previously derived multi-modal ELBO, we approximate the intractable prior $p(\mathbf{z_c|x})$ with $q(\mathbf{z_c|x,y})$.

\subsection{Cross-Modal Training Loss.}
Based on Eq. \ref{eq:cm_1} and Eq. \ref{eq:cm_2}, we derive the following loss function.
\begin{equation}
    \begin{aligned}
    &\mathcal{L}_{\textrm{CM}} = - \mathbb{E}_{q(\mathbf{z_m^x|u}) q(\mathbf{z_c|y})}[\log p(\mathbf{x|z_m^x,z_c})] \\
    &+ KL[q(\mathbf{z_m^x|u})||p(\mathbf{z_m^x|u})] + KL[q(\mathbf{z_c|y})||p(\mathbf{z_c|y})] \\
    &-\mathbb{E}_{q(\mathbf{z_c|x}) p(\mathbf{z_m^y})}[\log p(\mathbf{y|z_c,z_m^y})] + KL[q(\mathbf{z_c|x})||p(\mathbf{z_c|x})]
\end{aligned}
\end{equation}
Note that the distributions $p(\mathbf{z_m^x|u})$, $p(\mathbf{z_c|y})$, and $p(\mathbf{z_c|x})$ are unknown and intractable. We therefore approximate them using their respective posterior distributions from multi-modal modeling, namely $q(\mathbf{z_m^x|x,u})$, $q(\mathbf{z_c|x,y})$ and $q(\mathbf{z_c|x,y})$. 

To standardize notation, we reformulate $\mathcal{L}_{\textrm{CM}}$. In this formulation, all distributions denoted by $p$ maintain their original meaning but use different notation, except for $p(\mathbf{x|z_m^x}, \mathbf{z_c})$ and $p(\mathbf{y|z_c}, \mathbf{z_m^y})$, which remain consistent with multi-modal modeling. Aligning these notations with the prior distributions from multi-modal modeling yields:
\begin{equation}
    \begin{aligned}
        \mathcal{L}_{\textrm{CM}} &= - \mathbb{E}_{p(\mathbf{z_m^x|u}) p(\mathbf{z_c|y})}[\log p(\mathbf{x|z_m^x,z_c})] \\
        &+ KL[p(\mathbf{z_m^x|u})||q(\mathbf{z_m^x|x,u})] 
        + KL[p(\mathbf{z_c|y})||q(\mathbf{z_c|x,y})] \\
        &-\mathbb{E}_{p(\mathbf{z_c|x}) p(\mathbf{z_m^y})}[\log p(\mathbf{y|z_c,z_m^y})] \\
        &+ KL[p(\mathbf{z_c|x})||q(\mathbf{z_c|x,y})]
    \end{aligned}
\end{equation}

By now, we have derived two ELBOs for both multi-modal and cross-modal modeling, as well as the corresponding loss functions based on them.

\section{Analysis of Spillover Effect in Latent Disentanglement of miVAE}
We explain how miVAE avoids the \textit{spillover} effect and extracts the most relevant latent variables for neural activity and visual stimuli, with a primary focus on neural activity $\mathbf{x}$.

According to our modeling, $\mathbf{u} \in \mathbb{R}^{N\times3}$ represents 3D coordinates, and $\mathbf{y} \in \mathbb{R}^{H\times W\times T}$ represents the visual stimulus. These come from two different spaces and represent fundamentally different types of information. Thus, we assume $\mathbf{u}$ and $\mathbf{y}$ are statistically independent. Based on this, \(\mathbf{z_m^x \perp y | (x, u)}\) and \(\mathbf{z_c^x \perp u | (x, y)}\), we have:
\begin{equation}
    \begin{cases}
        q(\mathbf{z_m^x | x, u, y}) = q(\mathbf{z_m^x | x, u})\\
        q(\mathbf{z_c^x | x, u, y}) = q(\mathbf{z_c^x | x, y})
    \end{cases}
\end{equation}

Furthermore, we use the introduced variational distribution $q(\mathbf{z_m^x, z_c^x| x, u, y})$ to approximate $p(\mathbf{z_m^x, z_c^x| u, y}) = p(\mathbf{z_m^x | u})p(\mathbf{z_c | y})$. The optimization process involves an independent assumption:
\begin{equation}
    q(\mathbf{z_m^x, z_c^x| x, u, y}) = q(\mathbf{z_m^x | x, u}) q(\mathbf{z_c^x | x, y})
\end{equation}

To demonstrate that miVAE can avoid the spillover effect, we will analyze it from the perspectives of (1) \textbf{Loss function} and (2) \textbf{mutual information}:

(1) \textbf{Loss function}. The miVAE loss function includes two KL divergence terms, $ KL[q(\mathbf{z_m^x|x,u})||p(\mathbf{z_m^x|u})]$ and $KL[q(\mathbf{z_c|x,y})||p(\mathbf{z_c|y})]$. These KL divergence terms enforce that $\mathbf{z_m^x}$ and $\mathbf{z_c^x}$ follow their prior distributions $p(\mathbf{z_m^x|u})$ and $p(\mathbf{z_c|y})$, respectively, ensuring the disentanglement of latent variables and avoiding the spillover effect. 

(2) \textbf{Mutual information}. 
For mutual information $I(\mathbf{z_m^x; z_c^x|x,u,y})$, we have:
\begin{equation}
\resizebox{\linewidth}{!}{$
    \begin{aligned}
        I(\mathbf{z_m^x; z_c^x|x,u,y}) &= \int\int q(\mathbf{z_m^x, z_c^x|x,u,y}) \log \frac{q(\mathbf{z_m^x, z_c^x|x,u,y})}{q(\mathbf{z_m^x|x,u,y}) q(\mathbf{z_c^x|x,u,y})} d\mathbf{z_m^x} d\mathbf{z_c^x} \\
        &= \int\int q(\mathbf{z_m^x, z_c^x|x,u,y}) \log \frac{q(\mathbf{z_m^x|x,u}) q(\mathbf{z_c^x|x,y})}{q(\mathbf{z_m^x|x,u}) q(\mathbf{z_c^x|x,y})} d\mathbf{z_m^x} d\mathbf{z_c^x} \\
        &=0
    \end{aligned}
$}
\end{equation}
This shows that the mutual information between $\mathbf{z_m^x}$ and $\mathbf{z_c}$ given $\mathbf{x,u,y}$ is zero.

Regarding the visual stimulus analysis $\mathbf{y}$, our model employs a single conditional prior for $\mathbf{y}$, namely the neural activity $\mathbf{x} \in \mathbb{R}^{N\times T}$. To model the most correlated variables $\mathbf{z_c}$ between $\mathbf{x}$ and $\mathbf{y}$, we introduce an additional latent variable $\mathbf{z_m^y}$ with prior distribution $\mathcal{N}(\mathbf{0,1})$ to capture uncorrelated components. This leads to:
\begin{equation}
    q(\mathbf{z_m^y, z_c^y| x, y}) = q(\mathbf{z_m^y | y}) q(\mathbf{z_c^y| x, y})
\end{equation}

The spillover effect mitigation mechanism can be further analyzed through the (1) \textbf{Loss function} and (2) \textbf{mutual information}:

(1) \textbf{Loss function}. The loss function includes two essential KL divergence terms: $ KL[q(\mathbf{z_m^y|y})||p(\mathbf{z_m^y})]$ and $KL[q(\mathbf{z_c|x,y})||p(\mathbf{z_c|x})]$. These terms constrain $\mathbf{z_m^y}$ and $\mathbf{z_c^y}$ to follow their respective prior distributions $p(\mathbf{z_m^y})\sim \mathcal{N}(\mathbf{0,1})$ and $p(\mathbf{z_c|x})$, ensuring effective latent variable disentanglement.

(2) \textbf{Mutual information}. For mutual information $I(\mathbf{z_m^y; z_c^y|x,y})$, we have:
\begin{equation}
\resizebox{\linewidth}{!}{$
    \begin{aligned}
        I(\mathbf{z_m^y; z_c^y|x,y}) &= \int\int q(\mathbf{z_m^y, z_c^y|x,y}) \log \frac{q(\mathbf{z_m^y, z_c^y|x,y})}{q(\mathbf{z_m^y|y}) q(\mathbf{z_c^y|x,y})} d\mathbf{z_m^y} d\mathbf{z_c^y} \\
        &= \int\int q(\mathbf{z_m^y, z_c^y|x,y}) \log \frac{q(\mathbf{z_m^y|y}) q(\mathbf{z_c^y|x,y})}{q(\mathbf{z_m^y|y}) q(\mathbf{z_c^y|x,y})} d\mathbf{z_m^y} d\mathbf{z_c^y} \\
        &=0
    \end{aligned}
$}
\end{equation}
This shows that the mutual information between $\mathbf{z_m^y}$ and $\mathbf{z_c^y}$ given $\mathbf{x,u,y}$ is zero.

These theoretical analyses demonstrate that miVAE effectively facilitates the exploration of maximally correlated latent variables between neural activity and visual stimuli, establishing a foundation for advanced latent encoding and decoding modeling.

Experimental validation was conducted on three unseen mice from the validation set. We evaluated the correlation between $\mathbf{z_m^x}$ and $\mathbf{z_c^x}$ for calcium signals, and between $\mathbf{z_m^y}$ and $\mathbf{z_c^y}$ for visual stimuli. Table \ref{table:corr} presents these results. The calcium signals, supported by comprehensive prior information, exhibit minimal correlation between latent variables. The visual stimuli analysis reveals some residual correlation between $\mathbf{z_m^y}$ and $\mathbf{z_c^y}$, attributable to the limited prior information for $\mathbf{z_m^y}$. Enhanced prior information could potentially reduce this correlation further.

\begin{table}[ht]
    \centering
    \begin{tabular}{lc}
        \toprule
        Modal              & Correlation \\
        \midrule
        Calcium signal ($\mathbf{z_m^x}$, $\mathbf{z_c^x}$)    &0.0255 \\
        Visual stimulus ($\mathbf{z_m^y}$, $\mathbf{z_c^y}$)     &0.3044 \\
        \bottomrule
    \end{tabular}
\caption{\textbf{Correlation Analysis of Latent Variables.} Pearson correlation coefficients between individual-specific ($\mathbf{z_m^x}$) and shared ($\mathbf{z_c^x}$) latent variables for calcium signals, and between stimulus-specific ($\mathbf{z_m^y}$) and shared ($\mathbf{z_c^y}$) latent variables for visual stimuli. Results are computed for three new mice from the validation set.}
\label{table:corr}
\end{table}

\section{Detailed Architecture of miVAE}

In this section, we provide a comprehensive description of the network architectures used in our study, specifically $\boldsymbol{\lambda}$, and $\boldsymbol{\theta, \phi, \vartheta, \psi}$. The module $\boldsymbol{\phi}$ encodes the calcium signal $\mathbf{x}$ within the miVAE framework, facilitating the approximate inference of the latent variables $\mathbf{z_m^x}$ and $\mathbf{z_p}$ of $\mathbf{x}$. The function $\boldsymbol{\theta}$ reconstructs the original calcium signals $\mathbf{x}$ based on $\mathbf{z_m^x}$ and $\mathbf{z_p}$. The network $\boldsymbol{\lambda}$ derives the individual prior distribution $p(\mathbf{z_m^x|\mathbf{u}})$, relying on individual-specific neural coordinate information. The module $\boldsymbol{\psi}$ encodes the visual stimulus $\mathbf{y}$ within the miVAE framework, facilitating the approximate inference of the latent variables $\mathbf{z_m^y}$ and $\mathbf{z_p}$ of $\mathbf{y}$. The function $\boldsymbol{\vartheta}$ reconstructs the video $\mathbf{y}$ based on $\mathbf{z_m^y}$ and $\mathbf{z_p}$. The detailed computation pipeline are as follows:

\textbf{The architecture of $\boldsymbol{\phi, \theta}$ for Calcium signal}.

For any given calcium signal $\mathbf{x} \in \mathbb{R}^{1 \times N \times T}$, where $N$ represents the number of neurons and $T$ represents the time window, we first applied a 2D convolution to map it into a feature space $\mathbf{x_{feat}} \in \mathbb{R}^{c \times N \times T}$. The signal was then adaptively downsampled to $\mathbf{x_{feat}} \in \mathbb{R}^{c' \times N' \times T}$, with $c'=64, N'=1024$. This was followed by two 2D residual blocks to enhance the features. Thus, neural activity inputs from different subjects could be uniformly represented as $\mathbf{x_{feat}}$.

Next, we used the encoder from the UNet\footnote{\url{https://github.com/hojonathanho/diffusion/blob/master/diffusion_tf/models/unet.py}} \cite{ho2020denoising} to further compress and encode $\mathbf{x_{feat}}$ into a more compact representation $\mathbf{x_{feat}} \in \mathbb{R}^{c \times n \times T}$, where $n=\frac{N'}{8}$ and $t=\frac{T}{8}$. This encoded feature was then used to generate the distributions for $\mathbf{z_m^x}$ and $\mathbf{z_p}$. For $\mathbf{z_m^x}$, $\mathbf{x_{feat}}$ passed through two 2D residual blocks to produce $\mathbf{{z_m^x}^{mean}} \in \mathbb{R}^{c \times n \times T}$ and $\mathbf{{z_m^x}^{log\_var}} \in \mathbb{R}^{c \times n \times T}$. Similarly, for $\mathbf{z_p} \in \mathbb{R}^{c \times n \times T}$, $\mathbf{x_{feat}}$ was processed through two more 2D residual blocks to generate $\mathbf{{z_p}^{mean}}$ and $\mathbf{{z_p}^{log\_var}}$.

After obtaining $\mathbf{z_m^x}$ and $\mathbf{z_p}$ through sampling, we concatenated them to form a bi-identifiable latent variable $\mathbf{z} \in \mathbb{R}^{2c \times n \times T}$. We then used the UNet decoder \cite{ho2020denoising} to generate the features of the anticipated calcium signal, represented as $\mathbf{x_{feat}}' \in \mathbb{R}^{c' \times N' \times T}$. These recovered features were upsampled to match the original dimensions of $\mathbf{x}$ using bicubic interpolation, resulting in $\mathbf{x_{feat}}' \in \mathbb{R}^{c' \times N \times T}$. Two residual blocks were then used to further refine these features. Finally, two convolutional layers reduced the feature dimensions to 1, producing the reconstructed signal $\mathbf{x}' \in \mathbb{R}^{1 \times N \times T}$.

\textbf{The architecture of 3d position encoder $\boldsymbol{\lambda}$}.

For the neural coordinate information $\mathbf{u} \in \mathbb{R}^{3 \times N}$ of any individual, we first project it into a feature space $\mathbf{u_{feat}} \in \mathbb{R}^{c \times N}$ using a one-dimensional convolution, where $c=16$. We then downsample $\mathbf{u_{feat}}$ to match the neuron dimension of $\mathbf{z_c}$ with a one-dimensional adaptive pooling module, obtaining $\mathbf{u_{feat}} \in \mathbb{R}^{c \times d}$. Next, $\mathbf{u_{feat}}$ is refined using three one-dimensional residual blocks. A final one-dimensional convolution extracts the desired features, which are then repeated to align with the temporal dimension of $\mathbf{z_c}$, resulting in $\mathbf{u_{feat}} \in \mathbb{R}^{c \times d \times T}$. Finally, two two-dimensional residual blocks output the mean $\mathbf{u_{mean}} \in \mathbb{R}^{c \times d \times T}$ and log-variance $\mathbf{u_{log\_var}} \in \mathbb{R}^{c \times d \times T}$ of the prior $\mathbf{u}$.

\textbf{The architecture of $\boldsymbol{\psi, \vartheta}$ for Visual Stimulus}.

We adopted the 'core' component from the first-place solution\footnote{\url{https://github.com/lRomul/sensorium}} of the Sensorium 2023 Competition \cite{turishcheva2023dynamic}. This solution features an efficient 3D video encoder, which replaces conventional 3D depth-wise convolution with spatial 2D depth-wise convolution followed by temporal 1D depth-wise convolution. It also incorporates the Squeeze-and-Excitation operation to enhance focus on significant temporal and spatial information. The encoder $\boldsymbol{\phi}$ and decoder $\boldsymbol{\theta}$ have similar architectures, except for down-sampling and up-sampling layers.

To tailor this approach, we modified the input channels to 1. For any dynamic visual stimulus $\mathbf{y} \in \mathbb{R}^{1 \times T \times h \times w}$, the encoder generates the feature representation $\mathbf{y_{feat}} \in \mathbb{R}^{c \times T \times h' \times w'}$. This is reshaped to $\mathbf{y_{feat}} \in \mathbb{R}^{c \times n \times T}$ and passed through four 2D residual blocks, producing the mean $\mathbf{{z_m^y}_{mean}} \in \mathbb{R}^{c \times d/2 \times T}$ and log-variance $\mathbf{{z_m^y}_{log\_var}} \in \mathbb{R}^{c \times d/2 \times T}$ of $\mathbf{z_m^y}$, and the mean $\mathbf{{z_c}_{mean}} \in \mathbb{R}^{c \times d/2 \times T}$ and log-variance $\mathbf{{z_c}_{log\_var}} \in \mathbb{R}^{c \times d/2 \times T}$ of $\mathbf{z_c}$. 

Next, we reconstruct the visual stimulus from $\mathbf{z_m^y}$ and $\mathbf{z_c}$. They are mapped to feature space $\mathbb{R}^{c \times T \times d/2}$ via two 2D residual blocks, combined, and reshaped to $\mathbb{R}^{c \times T \times h' \times w'}$. Using an architecture similar to the encoder $\boldsymbol{\psi}$, but with up-sampling layers replacing down-sampling ones, the decoder $\boldsymbol{\vartheta}$ reconstructs the visual stimulus $\mathbf{y} \in \mathbb{R}^{1 \times T \times h \times w}$.

\subsection{Detailed Architecture of Latent Coding Networks}
In this section, we describe the Linear and non-Linear network architectures used for latent space encoding and decoding. It is important to note that the structures for encoding and decoding are consistent, whether linear or non-linear, with the difference lying in the learning objectives.

\textbf{Linear coding networks.} 

For linear coding networks, we use linear layers for computation. Using the encoding process as an example, for the input latent variable $\mathbf{z_c^y} \in \mathbb{R}^{c \times d/2 \times T}$ of neural activity $\mathbf{x}$, we use two linear layers, with parameters $\mathbf{W_e^{mean}}$, $\mathbf{b_e^{mean}}$, $\mathbf{W_e^{log\_var}}$ and $\mathbf{b_e^{log\_var}}$, to learn the latent variable $\mathbf{z_c^x} \in \mathbb{R}^{c \times d/2 \times T}$ from the visual stimulus. Specifically, we use distribution matching strategy for this mapping:
\begin{equation}
    \begin{cases*}
    \mathbf{mean}_\mathbf{z_c^x} = \mathbf{W_e^{mean}}^{T} * \mathbf{mean}_\mathbf{z_c^y} + \mathbf{b_e^{mean}} \\
    \mathbf{log\_var}_\mathbf{z_c^x} = \mathbf{W_e^{log\_var}}^{T} * \mathbf{log\_var}_\mathbf{z_c^y} + \mathbf{b_e^{log\_var}}
    \end{cases*}
\end{equation}
Similarly, for the decoding model, we have:
\begin{equation}
    \begin{cases*}
    \mathbf{mean}_\mathbf{z_c^y} = \mathbf{W_e^{mean}}^{T} * \mathbf{mean}_\mathbf{z_c^x} + \mathbf{b_e^{mean}} \\
    \mathbf{log\_var}_\mathbf{z_c^y} = \mathbf{W_e^{log\_var}}^{T} * \mathbf{log\_var}_\mathbf{z_c^x} + \mathbf{b_e^{log\_var}}
    \end{cases*}
\end{equation}
where $\mathbf{W_d^{mean}}$, $\mathbf{b_d^{mean}}$, $\mathbf{W_d^{log\_var}}$ and $\mathbf{b_d^{log\_var}}$ are the learnable parameters of decoding models, sharing the same shape as $\mathbf{W_e^{mean}}$, $\mathbf{b_e^{mean}}$, $\mathbf{W_e^{log\_var}}$ and $\mathbf{b_e^{log\_var}}$.

\textbf{Non-linear coding networks.}

For nonlinear coding networks, we use more complex neural networks. Similar to the description of linear coding, we use the encoding model as an example. The input latent variable $\mathbf{z_c^y} \in \mathbb{R}^{c \times d/2 \times T}$ of neural activity $\mathbf{x}$ is mapped to the target latent variable $\mathbf{z_c^x} \in \mathbb{R}^{c \times d/2 \times T}$ of visual stimulus $\mathbf{y}$. Specifically, this mapping process is based on 2 sub-networks with three 2D convolution blocks, denoted as $\mathbf{g^{e}_{mean}}$ and $\mathbf{g^{e}_{log\_var}}$. These blocks consist of residual blocks from UNet\footnote{\url{https://github.com/hojonathanho/diffusion/blob/master/diffusion_tf/models/unet.py}} \cite{ho2020denoising}, which map the Gaussian distribution parameters $\mathbf{mean}_{\mathbf{z_c^y}}$ and $\mathbf{log\_var}_{\mathbf{z_c^y}}$ to $\mathbf{mean}_{\mathbf{z_c^x}}$ and $\mathbf{log\_var}_{\mathbf{z_c^x}}$. Specifically:
\begin{equation}
    \begin{cases*}
    \mathbf{mean}_\mathbf{z_c^x} = \mathbf{g^{e}_{mean}}(\mathbf{mean}_\mathbf{z_c^y}) \\
    \mathbf{log\_var}_\mathbf{z_c^x} = \mathbf{g^{e}_{log\_var}}(\mathbf{log\_var}_\mathbf{z_c^y})
    \end{cases*}
\end{equation}
Similarly, for the decoding model, we have:
\begin{equation}
    \begin{cases*}
    \mathbf{mean}_\mathbf{z_c^x} = \mathbf{g^{d}_{mean}}(\mathbf{mean}_\mathbf{z_c^y}) \\
    \mathbf{log\_var}_\mathbf{z_c^x} = \mathbf{g^{d}_{log\_var}}(\mathbf{log\_var}_\mathbf{z_c^y})
    \end{cases*}
\end{equation}
where $\mathbf{g^{d}_{mean}}$ and $\mathbf{g^{d}_{log\_var}}$ are the decoding networks, with identical architecture as $\mathbf{g^{e}_{mean}}$ and $\mathbf{g^{e}_{log\_var}}$.

\section{Experimental Details}
\subsection{V1 Dataset}
\label{SI:dataset-visualcortex}

While the Allen Brain Observatory \cite{de2020large} represents a comprehensive neural dataset with approximately 60,000 neurons recorded from 243 mice across multiple visual areas (V1, LM, AL, PM, AM, and RL), its limitations for our specific research objectives warrant consideration. The relatively low number of effective neurons per individual mouse in V1 introduces potential biases that could impact modeling accuracy, as discussed in the Introduction. Additionally, despite the variety in visual stimulus types, the dataset contains only three video stimuli shared across all mice, potentially constraining the model's ability to capture the full complexity of visual-neural relationships and increasing the risk of overfitting.

The Sensorium dataset \cite{turishcheva2023dynamic} better aligns with our research requirements by providing dense neuronal sampling (over 7,000 neurons per mouse) and enhanced stimulus diversity, with minimal overlap in visual stimuli between subjects. These characteristics make it particularly suitable for our data-driven approach to modeling visual processing. Our analysis utilizes this publicly available mouse V1 dataset from the SENSORIUM 2023 competition, which encompasses neural recordings from 10 mice exposed to dynamic visual stimuli, comprising approximately 700 minutes of neural activity from 78,853 neurons. Looking ahead, we aim to extend this research through the collection of more diverse datasets, with a particular focus on non-human primates, to further advance our understanding of V1 mechanisms.

Therefor, we utilized the publicly available mouse V1 dataset from the SENSORIUM 2023\footnote{\url{https://www.sensorium-competition.net}}\footnote{\url{https://gin.g-node.org/pollytur/Sensorium2023Data}}\footnote{\url{https://gin.g-node.org/pollytur/sensorium_2023_dataset}} competition. This dataset includes data from 10 mice exposed to dynamic visual stimuli, totaling approximately 700 minutes of recorded neural activity from 78,853 neurons.

\textbf{Visual Stimuli.} The dynamic visual stimuli were sourced from cinematic movies and the Sports-1M dataset \cite{karpathy2014large, turishcheva2023dynamic, wang2023towards}. Each stimulus was converted to grayscale and presented at a frame rate of 30 Hz for durations between 8 and 11 seconds. The movies were displayed on a 31.8 by 56.5 cm monitor positioned 15 cm from and perpendicular to the left eye of the mice. The movies were spatially downsampled to 36 by 64 pixels, corresponding to a resolution of 3.4$^{\circ}$/pixel at the center of the screen. Notably, only pairs of mice were presented with the same visual stimuli.

\textbf{Neural Responses and 3D Coordinates.} Neural responses were recorded using two-photon calcium imaging \cite{sofroniew2016large} in head-fixed mice. The calcium signals were deconvolved to infer spike trains \cite{turishcheva2023dynamic}. The initial sampling rate of 8 Hz was resampled to 30 Hz to synchronize with the visual stimulus frame rate. Each neuron's 3D coordinates were also recorded.

For each mouse, the dataset consists of 60 minutes of recordings for training and 10 minutes for validation, covering all visual stimuli and corresponding neural responses. Each validation set includes repeated stimuli (8-10 repeats) and their corresponding calcium signals to ensure consistency. Thus, the complete dataset for 10 mice amounts to 700 minutes of recordings (10 mice × 70 minutes each). In our experiments, we deviated from this setup by using data from 7 mice for training and data from 3 mice for validation. Although the dataset includes behavioral information, this aspect was not utilized in our experiments. For further details about the dataset, please refer to \cite{turishcheva2023dynamic, wang2023towards}.

\subsection{Dataset split} 
\label{SI:dataset_split_V1}

\begin{table*}[t]
    \centering
    \caption{\textbf{Comprehensive Dataset Organization and Video Stimulus Pairing Information.} The dataset comprises recordings from 10 mice, split into training (7 mice) and validation (3 mice) sets. Each mouse is assigned a unique identifier and is paired with another mouse that viewed identical video stimuli, enabling cross-validation of neural responses to the same visual input. The pairings are designed to include mice from both training and validation sets, allowing for robust evaluation of model generalization across different subjects.}
    \begin{tabular}{cllc}
        \toprule
        \textbf{Split} & \textbf{Mouse ID} & \textbf{Dataset Identifier} & \textbf{Paired Mouse} \\
        \midrule
        \multirow{7}{*}{Training} 
        & Mouse 1 & dynamic29156-11-10-Video-8744edeac3b4d1ce16b680916b5267ce & Mouse 5 \\
        & Mouse 2 & dynamic29234-6-9-Video-8744edeac3b4d1ce16b680916b5267ce & Mouse 6 \\
        & Mouse 3 & dynamic29513-3-5-Video-8744edeac3b4d1ce16b680916b5267ce & Mouse 7 \\
        & Mouse 4 & dynamic29514-2-9-Video-8744edeac3b4d1ce16b680916b5267ce & Mouse 9 \\
        & Mouse 5 & dynamic29515-10-12-Video-9b4f6a1a067fe51e15306b9628efea20 & Mouse 1 \\
        & Mouse 6 & dynamic29623-4-9-Video-9b4f6a1a067fe51e15306b9628efea20 & Mouse 2 \\
        & Mouse 7 & dynamic29647-19-8-Video-9b4f6a1a067fe51e15306b9628efea20 & Mouse 3 \\
        \midrule
        \multirow{3}{*}{Validation} 
        & Mouse 8 & dynamic29228-2-10-Video-8744edeac3b4d1ce16b680916b5267ce & Mouse 10 \\
        & Mouse 9 & dynamic29712-5-9-Video-9b4f6a1a067fe51e15306b9628efea20 & Mouse 4 \\
        & Mouse 10 & dynamic29755-2-8-Video-9b4f6a1a067fe51e15306b9628efea20 & Mouse 8 \\
        \bottomrule
    \end{tabular}
    \label{SI:data_pair_info}
\end{table*}

In this specific V1 dataset, only two mice were required to watch the same video stimuli. The paired information is showing in Table \ref{SI:data_pair_info}.
\subsection{Training details} 
For all baseline models and miVAE, we use the AdamW optimizer \cite{loshchilov2017decoupled} with hyperparameters $(\beta_{1}, \beta_{2})=(0.9, 0.999)$ and a weight decay of 0.05. The batch size is 32, with each batch containing data from 16 frames. The peak learning rate is $1 \times 10^{-4}$, with a linear warm-up phase for the first 600 iterations. This rate is maintained for 200 epochs, then decays to $1 \times 10^{-6}$ using a cosine annealing strategy over the final 200 epochs. Training requires approximately 20 hours on 8 NVIDIA A100 GPUs (40G).

For all linear and non-linear encoding and decoding experiments, we also use the AdamW optimizer \cite{loshchilov2017decoupled} with the same hyperparameters $(\beta_{1}, \beta_{2})=(0.9, 0.999)$ and a weight decay of 0.05. The batch size remains 32, with each batch containing data from 16 frames. The peak learning rate is set to $2 \times 10^{-4}$, maintained for 100 epochs before decaying to $1 \times 10^{-6}$ using a cosine annealing strategy over the final 100 epochs. Training for these experiments requires approximately 4-6 hours on 8 NVIDIA A100 GPUs (40G).

\subsection{Implementation of Attribution Analysis} 
\textbf{Notations.} We denote the neural activity encoder as $\boldsymbol{\phi}$ and the neural activity data as $\mathbf{x}$. Similarly, the visual stimulus encoder is represented by $\boldsymbol{\psi}$, with the visual stimulus data denoted as $\mathbf{y}$.

\textbf{Pipelines.} Our implementation leverages PyTorch for automatic differentiation. To illustrate our approach, we describe the process of attributing the latent variable $\mathbf{z_c^y}$ of the visual stimulus $\mathbf{y}$ to the neural activity $\mathbf{x}$.

Initially, we obtain $\mathbf{z_c^y}$ by applying the encoder $\boldsymbol{\psi}$ to the actual visual stimulus. This is followed by latent space encoding, represented as $f_{\text{e}}(\mathbf{z_c^y})$. To enable gradient computation, we generate a random neural activity input $\mathbf{x}^{fake}$, which is processed by $\boldsymbol{\phi}$ to create a computational graph and corresponding pseudo-latent variable $\mathbf{z_c}^{fake}$.

Subsequently, we copy the tensor values of $f_{\text{e}}(\mathbf{z_c^y})$ to $\mathbf{z_c}^{fake}$. Automatic differentiation is then applied to derive the attributed weights of the real neural activity $\mathbf{x}$.

This methodological framework is consistently applied for both neural activity and visual stimulus attribution processes.

\subsection{Stimulus Classification Based on Important Subpopulations} 
For the V1 stimulus classification, we validated the identified important neural subpopulations using data from three new mice in the validation set. The sub-dataset includes six different stimuli for single mouse, each repeated ten times. We split each stimulus dataset into a 70\% training set (42 samples) and a 30\% validation set (18 samples). We applied eight classical classifiers: Linear SVM, RBF SVM, Polynomial SVM, KNN (with n$\_$neighbors=2), Decision Tree, Random Forest, Naïve Bayes, and MLP\footnote{All these algorithms are included in the Python 'sklearn' package}. Each classifier was trained and validated separately. We reported the mean and variance of their performance in Figure \ref{fig:attribution_analysis}.D.



\end{document}